\documentclass[aps,twocolumn,prb]{revtex4}
\usepackage{graphicx}

\begin{document}
\title{\bf Quantum dot dephasing by fractional quantum Hall edge states}
\date{\today}
\author{T. K. T. Nguyen$^{1,2,3}$, A. Cr\'epieux$^{1,2}$, T. Jonckheere$^{1}$, A. V. Nguyen$^{3}$, Y. Levinson$^4$, and T. Martin$^{1,2}$}
\affiliation{$^1$ Centre de Physique Th\'eorique, Case 907 Luminy, 13288 Marseille cedex 9, France}
\affiliation{$^2$ Universit\'e de la M\'edit\'erann\'ee, 13288 Marseille Cedex 9, France}
\affiliation{$^3$ Institute of Physics and Electronics, 10 Dao Tan, Cong Vi, Ba Dinh, Hanoi, Vietnam}
\affiliation{$^4$ Department of Condensed Matter Physics, The Weizmann Institute of Science, Rehovot 76100, Israel}
%
%
\begin{abstract}
We consider the dephasing rate of an electron level in a quantum dot, placed next to a fluctuating edge current in the fractional quantum Hall effect. Using perturbation theory, we show that this rate has an anomalous dependence on the bias voltage applied to the neighboring quantum point contact, 
which originates from the Luttinger liquid physics which describes the Hall fluid. General expressions are obtained using a screened Coulomb interaction. 
The dephasing rate is strictly proportional to the zero frequency backscattering current noise, which allows 
to describe exactly the weak to strong backscattering crossover using the Bethe-Ansatz solution.   
\end{abstract}

\maketitle

Transport through a quantum dot is typically affected by the environment which surrounds it: the level of such a dot 
acquires a finite linewidth if this environment has strong charge fluctuations which couple to the dot.
Several seminal experiments, performed with a quantum dot embedded in an Aharonov-Bohm loop, probed the phase coherence of 
transport when this dot is coupled to a controlled environment, such as a quantum point contact (QPC) with a fluctuating current \cite{yacoby1,yacoby2,buks_PRL,schuster}. Charge fluctuations in the QPC create a fluctuating potential at the dot, modulate 
the electron levels in the dot, and destroy the coherence of the transmission through the dot \cite{buks_nature,sprinzak}.
The destruction of coherence is called ``dephasing''. A general theoretical framework for describing dephasing 
has been presented in Ref. \onlinecite{levinson_euro39,aleiner}, and was applied to a quantum Hall geometry   
\cite{levinson_PRB_61}, and  to a normal metal-superconductor QPC\cite{guyon_martin_lesovik}. 
In all the above, the dephasing rate typically 
increases when the voltage bias of the QPC is increased.

The purpose of the present paper is to discuss the case of dephasing from a QPC in the fractional quantum Hall effect (FQHE)
regime\cite{laughlin}. QPC transmission can then be described by tunneling between edge states\cite{wen_92}, 
the quantized analog of classical 
skipping orbits of electrons. In this strongly correlated electron regime, edge states represent 
collective excitations of the quantum Hall fluid: depending on the pinching of the QPC, it is either 
FQHE quasiparticles or electrons which tunnel. 
It is particularly interesting because the current-voltage and the noise 
characteristics deviate strongly from the case of normal conductors \cite{kane_PRL92,kane_PRL94,chamon_PRB95}: 
for the weak backscattering (BS) case, the current at zero temperature may increase when the voltage bias is lowered, 
while in the strong BS case the $I(V)$ is highly non linear. 
It is thus important to address the issue of dephasing from a Luttinger liquid. Here, we consider the case of simple Laughlin fractions, 
with filling factor $\nu=1/m$ ($m$ odd integer).  
As in Ref. \onlinecite{levinson_euro39}, the dephasing of a state in the dot is induced by its capacitive 
coupling to the biased QPC, assuming that the level modulation in the dot is a Gaussian process,
and neglecting back-action effects.

\begin{figure}[h]
\centerline{\includegraphics[width=6cm]{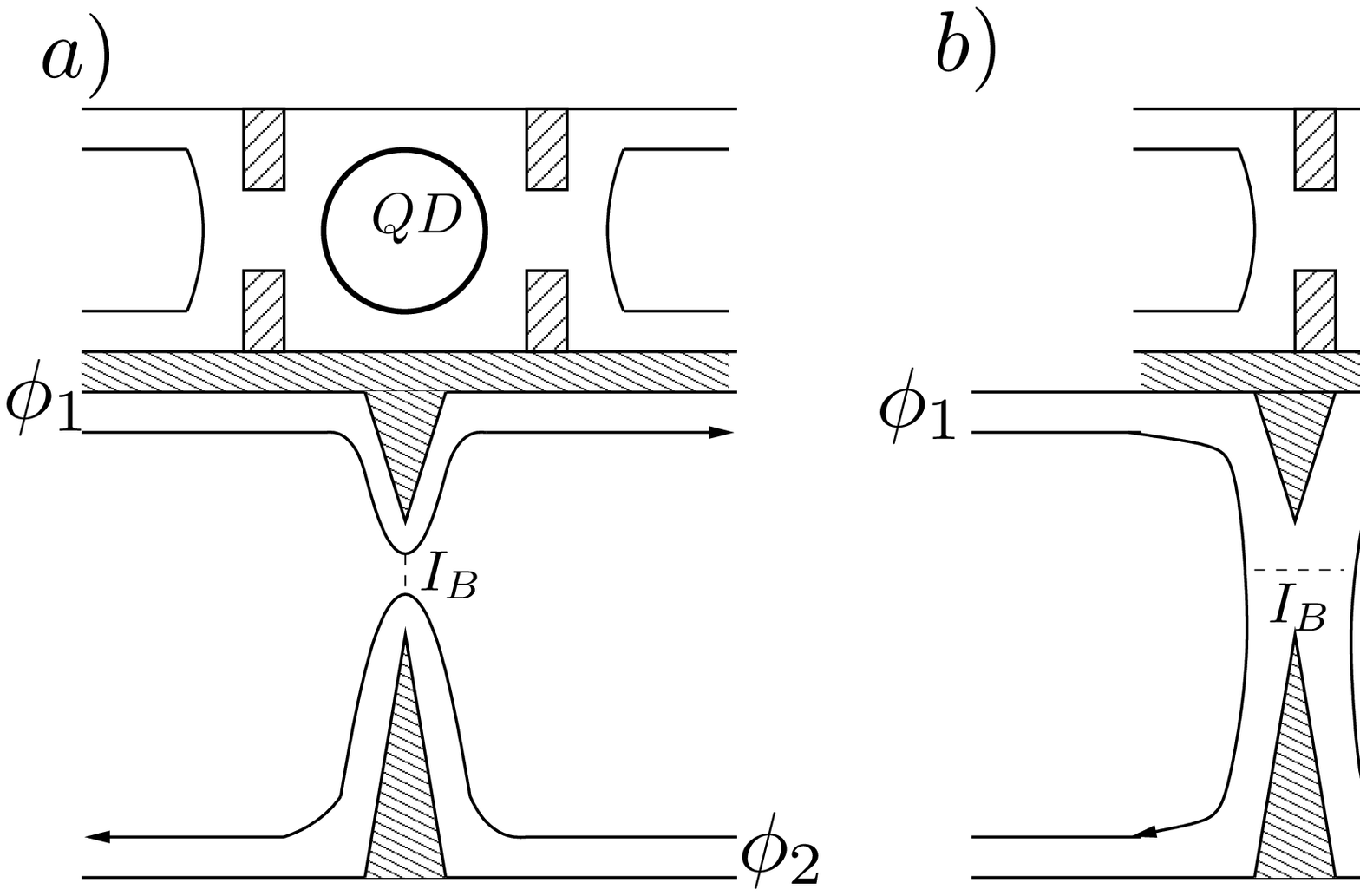}}
\caption{Schematic description of the setup: the quantum dot (top) is coupled capacitively to a quantum point contact in 
the FQHE regime: a) Case of weak backscattering, b) Case of strong BS.}
\label{fig1}
\end{figure}

In Fig. \ref{fig1}, 
a gate voltage controls the transmission in the fractional quantum Hall fluid through the QPC. 
The single level Hamiltonian for the dot
reads $H_{QD}=\epsilon_0 c^+c$, where $c^+$ creates an electron.
This dot is coupled capacitively to the nanostructure -- a point contact in the FQHE. 
The Hamiltonian which describes the edge modes in the absence of tunneling is: 
\begin{equation}
H_{0}=\frac{\hbar v_F}{4\pi}\int
dx[(\partial_x\phi_1)^2+(\partial_x\phi_2)^2]~,
\end{equation}
with $\phi_i(x)$ ($i=1,2$) is the Luttinger bosonic field, which relates to the electron density operator $\rho_i(x)$ by $\partial_x\phi_i(x)=\frac{\pi}{\sqrt{\nu}}\rho_i(x)$.

By varying the gate potential of QPC, one can switch from a weak BS situation, 
where the Hall liquid remains in one piece (Fig. \ref{fig1}a), to a strong BS situation where the Hall liquid is split in two
(Fig. \ref{fig1}b). In the former case, the entities which tunnel are edge quasiparticle excitations. In the latter case, 
between the two fluids, only electrons can tunnel. Here, we consider first the weak BS case, we use a 
duality transformation\cite{chamon_PRB96,kane_PRL92} to describe the strong BS case. 
The tunneling Hamiltonian between edges 1 and 2 reads:
\begin{equation}
H_t=e^{i\omega_0t}\Gamma_0 \psi^{+}_{2}(0)\psi_{1}(0)+h.c
\end{equation}
where we have used the Peierls substitution to include the voltage: for the weak BS, $\omega_0=e^\star V/\hbar$, ($e^\star=\nu e$ 
is the effective charge, $\nu$ is the filling factor), while $\omega_0=eV/\hbar$ for the strong BS case.
The quasiparticle operator in the case of weak BS is $\psi_i(x)=e^{i\sqrt{\nu}\phi_i(x)}/\sqrt{2\pi\alpha}$
(the spatial cutoff is $\alpha=v_F\tau_0$, with $\tau_0$ the temporal cutoff), 
and in the strong BS case the electron operator is obtained with the substitution 
$\nu\to 1/\nu$. 

The Hamiltonian describing the interaction between the dot and the QPC reads $H_{int}=c^{+}c\int dx f(x)\rho_1(x)$, 
with $f(x)$ is a Coulomb interaction kernel, which is assumed to include screening by the nearby gates $f(x)\simeq e^2 e^{-|x|/\lambda_s}/\sqrt{x^2+d^2}$, where $d$ is the distance from the dot to the edge, $\lambda_s$ is a screening length.

The dephasing of an electron state in a dot coupled to a fluctuating current is caused by the electron density fluctuations, 
which generate a fluctuating potential in the dot, resulting in a blurring of the energy level $\epsilon_0$. 
The dephasing rate, expressed in terms of irreducible charge fluctuations in the adjacent wire, is written as \cite{levinson_euro39,aleiner,levinson_PRB_61}: 
\begin{eqnarray}
\tau_{\varphi}^{-1}&=&\frac{1}{4\hbar^2}\int^{\infty}_{-\infty}\!\!\!\!\!\!dt\int\!\!\! dx f(x)\int\!\!\!
dx^\prime f(x^\prime)\nonumber\\
&&\times\langle\langle\rho_1(x,t)\rho_1(x^\prime,0)+\rho_1(x^\prime,0)\rho_1(x,t)\rangle\rangle~.
\end{eqnarray}
In normal and superconducting systems, the dephasing rate can be calculated using the scattering approach.
For Luttinger liquids and in particular for the FQHE, it is conveninent to use the Keldysh approach\cite{chamon_PRB95,martin_noise}.

Here a tunneling event (at $x=0$) creates an excitation which need to propagate to the location of the dot.
The equilibrium (zero point) contribution to the dephasing rate corresponds to the zero order in the tunneling amplitude $\Gamma_0$ (it is labeled $(\tau_{\varphi}^{-1})^{(0)}$).
There is no contribution to first order in the tunneling Hamiltonian, while the non-equilibrium contribution corresponds to the second order in $\Gamma_0$ exists, $\tau_{\varphi}^{-1}=(\tau_{\varphi}^{-1})^{(0)}+(\tau_{\varphi}^{-1})^{(2)}+...$. 
The dephasing rate in the weak BS case is obtained as\cite{guyon}: 
\begin{eqnarray}
(\tau_{\varphi}^{-1})^{(0)}&=&\frac{\nu}{4\pi^2\hbar^2}\int^{\infty}_{-\infty}\!\!\!\!\!\!dt \int\!\!\!dx f(x)\int\!\!\!dx^\prime f(x^\prime)\nonumber\\
&&
\times\sum_{\eta=\pm}\partial_{xx^\prime}^{2}G_{1}^{\eta-\eta}(x-x^\prime,t)~. 
\label{A0}
\end{eqnarray}
The bosonic Green's function is $ G_{i}^{\eta_1\eta_2}(x-x',t_1-t_2)=\langle \phi_{i}(x,t_{1}^{\eta_1})\phi_{i}(x',t_{2}^{\eta_2})-\phi_{i}^2\rangle$. 
The coefficients $\eta$,$\eta_{1,2}=\pm$ identify the upper/lower branch of the Keldysh contour.
For the $2^{nd}$ order, since $\psi_1,\psi_2$ are independent in the absence of tunneling, we obtain
\begin{eqnarray}
&&\!\!\!\!\!\!\!\!\!\!\!\sum_{\eta=\pm}\bigg \langle T_K\rho_1(x,t^\eta)\rho_1(x^\prime,t^{\prime-\eta})\frac{(-i)^2}{2\hbar^2}\int_K\!\!\!\! dt_1\int_K\!\!\!\!\! dt_2 H_t(t_1)H_t(t_2)\bigg \rangle\nonumber\\
&=&\!\!\!-\frac{\Gamma_0^2 \nu}{2\pi^2\hbar^2(2\pi\alpha)^2}\!\!\!\!\!\sum_{\eta,\eta_1,\eta_2,\epsilon_1,\epsilon_2}\!\!\!\!\!\int_{-\infty}^{\infty}\!\!\!\!\!\!\!dt_1\!\!\!\int_{-\infty}^{\infty}\!\!\!\!\!\!dt_2 e^{i(\epsilon_1\omega_0 t_1+\epsilon_2\omega_0t_2)}\eta_1\eta_2\nonumber\\
&&\!\!\!\!\!\!\!\!\!\times\! \langle T_K \partial_x \phi_1\!(x,t^\eta\!) \partial_{x^\prime} \phi_1(x^\prime\!,t^{\prime-\eta}\!) e^{i \sqrt{\nu}\epsilon_1\phi_1(0,t^{\eta_1}_{1})} e^{i \sqrt{\nu}\epsilon_2\phi_1(0,t^{\eta_2}_{2})}\! \rangle\nonumber\\
&&\!\!\!\!\!\!\!\!\!\times \langle T_K e^{-i \sqrt{\nu}\epsilon_1\phi_2(0,t^{\eta_1}_{1})} e^{-i \sqrt{\nu}\epsilon_2\phi_2(0,t^{\eta_2}_{2})}
\rangle~.
\label{A1}
\end{eqnarray}
Quasiparticle conservation imposes $\epsilon_1=-\epsilon_2\equiv\epsilon$, so
\begin{eqnarray}
(\tau_{\varphi}^{-1})^{(2)}\!\!&=&\!\!-\frac{\nu}{4\pi^2\hbar^4}\frac{\Gamma_0^2}{2(2\pi\alpha)^2}\!\!\int^{\infty}_{-\infty}\!\!\!\!\!\!dt \!\!\int\!\!\! dx f(x)\!\!\!\int\!\!\! dx^\prime f(x^\prime) \!\!\!\!\sum_{\eta,\eta_1,\eta_2,\epsilon}\!\!\!\!\eta_1\eta_2\nonumber\\
&&\!\!\!\!\!\!\!\!\!\!\!\!\!\!\!\!\!\!\!\!\!\!\!\!\!\!\!\!\!\!\!\!\times\!\!\int^{\infty}_{-\infty}\!\!\!\!\!\!dt_1\int^{\infty}_{-\infty}\!\!\!\!\!\!dt_2 e^{i\epsilon\omega_0(t_1-t_2)} e^{\nu G_{2}^{\eta_1\eta_2}(0,t_1-t_2)}e^{\nu G_{1}^{\eta_1\eta_2}(0,t_1-t_2)}\nonumber\\
&&\!\!\!\!\!\!\!\!\!\!\!\!\!\!\!\!\!\!\!\!\!\!\!\!\!\!\!\!\!\!\!\!\!\times\! \left\{\!\partial_{xx^\prime}^{2}G_{1}^{\eta-\eta}\!(x-x^\prime\!,t)+\nu[\partial_x G_{1}^{\eta\eta_1}\!(x,t-t_1)-\partial_x G_{1}^{\eta\eta_2}\!(x,t-t_2)]\right.\nonumber\\
&&\left.\!\!\!\!\!\!\!\!\!\!\!\!\!\!\!\!\!\!\!\!\times [\partial_{x^\prime}G_{1}^{-\eta\eta_1}(x^\prime,-t_1)-\partial_{x^\prime}G_{1}^{-\eta\eta_2}(x^\prime,-t_2)]\right\}~.
\label{A2}
\end{eqnarray}

The dephasing rate depends on the geometry of the set up via the length scales d, $\lambda_s$, and $\alpha$. 
The equivalent result for strong BS is obtained by replacing $\nu\rightarrow 1/\nu$ next to the Green's function (duality). 

The assumption of strong screening $\lambda_s\sim\alpha=v_F\tau_0$ is made ($f(x)\simeq 2 e^2\alpha\delta(x)/d$): it turns out that this assumption is not necessary, and it will be relaxed later on. 
Inserting the Green's function at finite temperature $G^{\eta\eta^\prime}_{1}(x,t)=-\ln\Big\{\sinh[\pi[(x/v_F-t)((\eta+\eta^\prime){\rm  sgn}(t)-(\eta-\eta^\prime))/2+i\tau_0]/\hbar\beta]\Big/\sinh[i\pi\tau_0/\hbar\beta]\Big\}$ in the dephasing rate (Eqs. (\ref{A0})-(\ref{A2})) is
$
(\tau_{\varphi}^{-1})^{(0)}=4e^4\tau_0^2\nu/(\pi\hbar^3\beta d^2)
$.
We perform the change of variables, $\tau=t_1-t_2$, $\tau_1=t-t_1$, and $\tau_2=t_2$ to obtain:
\hspace{-1.5cm}
\begin{eqnarray}
&&(\tau_{\varphi}^{-1})^{(2)}=-\frac{e^4\nu^2\Gamma_0^2}{4\hbar^6\beta^2\pi^2 v_F^2 d^2}\sum_\eta\int^{\infty}_{-\infty}\!\!\!\!\!\!d\tau \cos[\omega_0\tau]\nonumber\\
&&\times\Big[\frac{\sinh^{2\nu}(\frac{\pi}{\hbar\beta}i\tau_0)}{\sinh^{2\nu}[\frac{\pi}{\hbar\beta}(\eta\tau+i\tau_0)]}+\frac{\sinh^{2\nu}(\frac{\pi}{\hbar\beta}i\tau_0)}{\sinh^{2\nu}[\frac{\pi}{\hbar\beta}(-\eta\tau+i\tau_0)]}\Big]\nonumber\\
&&\times\int^{\infty}_{-\infty}\!\!\!\!\!d\tau_1 \left[{\rm sgn}(\tau_1)\coth[\frac{\pi}{\hbar\beta}(-\eta {\rm sgn}(\tau_1)\tau_1+i\tau_0)]\right.\nonumber\\
&&\left.+\coth[\frac{\pi}{\hbar\beta}(\eta\tau_1+i\tau_0)]\right]\nonumber\\
&&\times\int^{\infty}_{-\infty}\!\!\!\!\!d\tau_2\left[-{\rm sgn}(\tau_2)\coth[\frac{\pi}{\hbar\beta}(\eta {\rm sgn}(\tau_2)\tau_2+i\tau_0)]\right.\nonumber\\
&&\left.+\coth[\frac{\pi}{\hbar\beta}(\eta\tau_2+i\tau_0)]\right]~.
\end{eqnarray}
In the integral over $\tau$, we change variables to $t=-\tau\mp i\tau_0\pm i\hbar\beta/2$ for the first (second) term, and 
the integral now runs in the complex plane form $-\infty\mp i\tau_0\pm \hbar\beta/2$ to $+\infty\mp i\tau_0\pm \hbar\beta/2$. We 
bring it back to $(-\infty,+\infty)$ by deforming the contour because there are no poles in the integrand.
For $\tau_0\ll \omega_0^{-1},\hbar\beta$, one obtains
\begin{equation}
\hspace{-0.2cm}
(\tau_{\varphi}^{-1})^{\!(2)}\!=\frac{e^4\Gamma_0^2}{\pi^2\hbar^4 v_F^2 d^2}\frac{\nu^2\tau_0^{2\nu}}{\Gamma(2\nu)}\Big(\frac{2\pi}{\hbar\beta}\Big)^{\!2\nu-1}\!\!\!\!\!\!\!\cosh\!\Big(\frac{\omega_0\hbar\beta}{2}\Big)\!\Big|\Gamma(\nu+i\frac{\omega_0\hbar\beta}{2\pi}\!)\Big|^{\!2}~.
\label{gamma_nu}
\end{equation}
In the zero temperature limit, we have $(\tau_{\varphi}^{-1})^{(0)}=0$ and 
\begin{equation}
(\tau_{\varphi}^{-1})^{(2)}=\frac{e^4\Gamma_0^2}{\pi\hbar^4 v_F^2 d^2}\frac{\nu^2\tau_{0}^{2\nu}}{\Gamma(2\nu)}|\omega_0|^{2\nu-1}~.
\label{a2T0}
\end{equation}
Note that $(\tau_{\varphi}^{-1})^{(2)}=(e\tau_0/d)^2 S_I(0)$, with $S_I(0)=\int dt \langle\langle I(t)I(0)\rangle\rangle$ the zero frequency BS current noise.

\begin{figure}[h]
\centerline{\includegraphics[width=5.5cm, angle=-90]{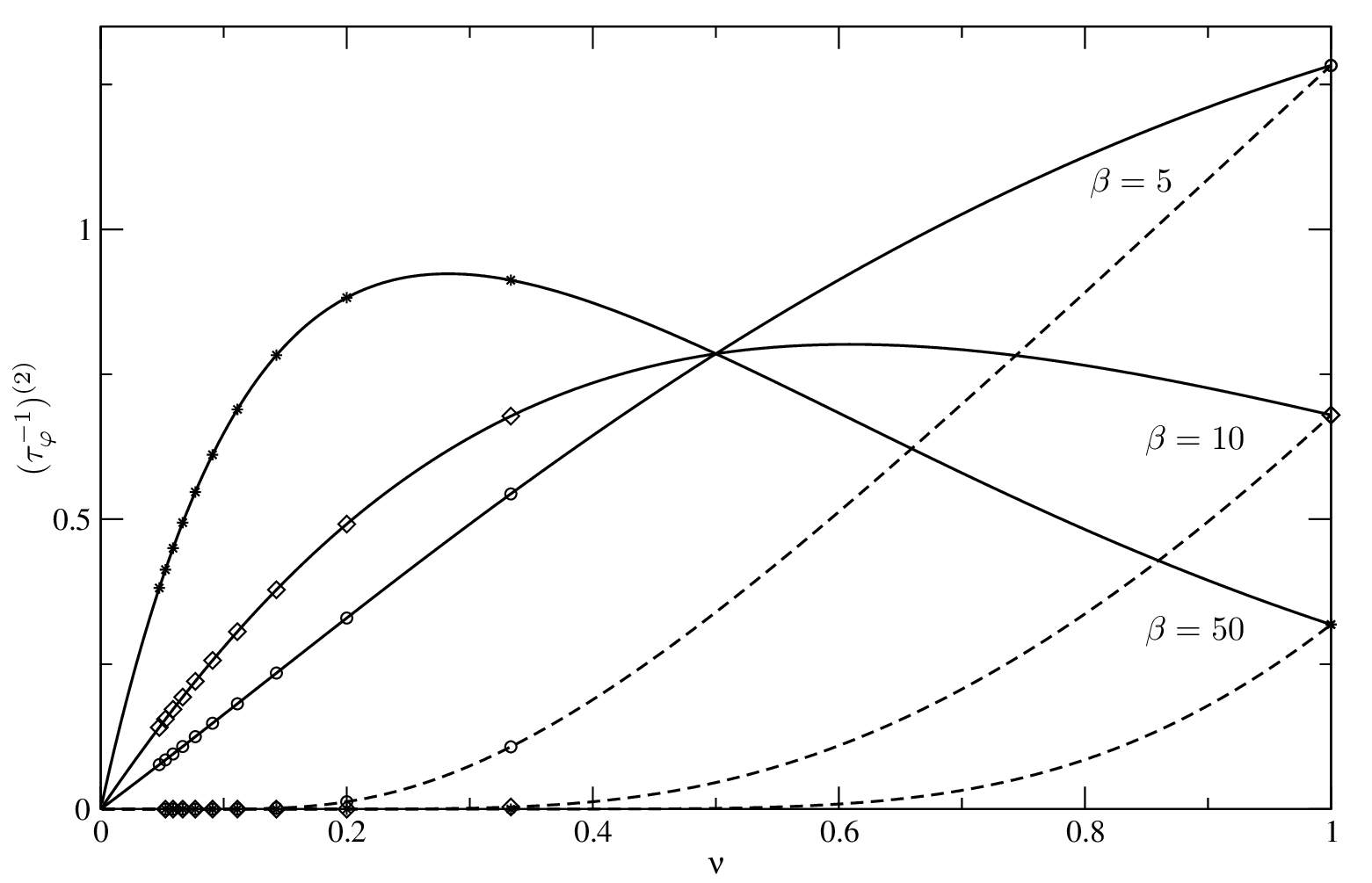}}
\caption{Dependance of the nonequilibrium contribution of the dephasing rate on the filling factor for both case weak (full line) and strong (dashed line)
backscattering at $\beta=5, 10, 50$ and QPC bias $eV=0.1$. 
The star, diamond and circle points correspond to the Laughlin fractions $\nu=1/m$, m odd integer.}
\label{fig2}
\end{figure}
The non-equilibrium contribution of the dephasing rate is proportional to the zero frequency noise in the quantum Hall liquid, which is computed 
in Refs. \onlinecite{chamon_PRB95,chamon_PRB96,kane_PRL94, martin_noise}. The theoretical predictions of noise in the weak and the strong BS limit have been verified in point contact experiments at filling factor $\nu=1/3,1/5$ \cite{sami,picciotto}. This is understood from the continuity equation,
which relates the current operator to the density operator\cite{guyon_martin_lesovik}. At zero temperature, the non-equilibrium dephasing rate of Eq. (\ref{a2T0}) for weak BS depends on the QPC bias with the exponent $2\nu-1<0$. This is in sharp contrast with Ref. \onlinecite{levinson_euro39}, where the QPC bias dependence is linear. We also calculate numerically this contribution at finite temperatures and consider it as a function of the filling factor or the QPC voltage bias. In our numerical calculations, we choose the inverse cutoff $\tau_{0}^{-1}$ as the energy scale, and the non-equilibrium contribution for the dephasing rate is plotted in units of $e^4\Gamma^2_0\tau_0/(\pi^2\hbar^4 v_F^2 d^2)$. 

In Fig. \ref{fig2}, we plot the dependence of this contribution on the filling factor $\nu$ for both weak and strong BS cases for several 
temperatures ($\beta=5, 10, 50$) at fixed QPC bias. $\nu$ is considered here as a continuous variable, while it has physical meaning only at Laughlin fractions\cite{laughlin}. For the strong BS case, the dephasing rate increases when the filling factor increases. At small $\nu$, it is zero, then, it increases rapidly. The higher the temperature, the faster the increase. For the weak BS case, the shape of the dephasing rate  depends on the ratio of QPC bias and temperature. At low temperature ($1/\beta\ll eV$), the dephasing rate function has a local maximum at $\nu<1/2$, the position of which depends on temperature: when the temperature increases, it gets closer to $\nu=1/2$ and its height decreases
The rate at $\nu=1$ is smaller than that at $\nu=1/3$. This result demonstrates that for two different filling factors, we can have comparable dephasing rates.
Around the crossover in temperature ($\beta eV\simeq1$), the local maximum in the dephasing rate broadens. At high temperature ($1/\beta > eV$), the dephasing rate increases when the filling factor increases. We find that the dephasing rates evaluated at different temperatures coincide at the (unphysical) value $\nu=1/2$, because the hyperbolic cosine multiplied by the squared modulus of the Gamma function in Eq. (\ref{gamma_nu}) does not depend on temperature, 
while at the same time the exponent $(2\nu -1)$ is zero: this is known for perturbative calculations of the backscattering current and noise.


\begin{figure}[t]
\centerline{\includegraphics[width=5.5cm, angle=-90]{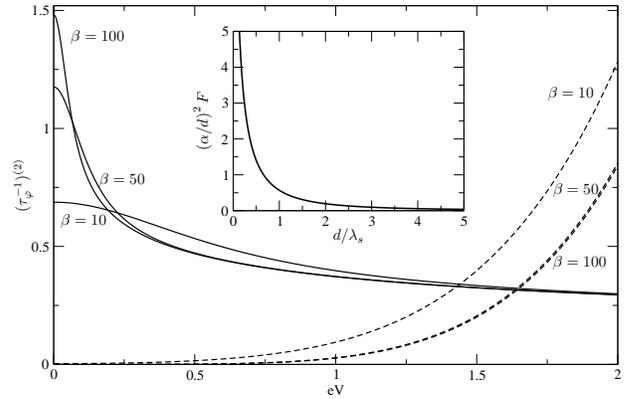}}
\caption{non-equilibrium contribution in the dephasing rate as a function of QPC bias with the filling factor $\nu=1/3$ at some values of temperature $\beta=10, 50, 100$ ($\beta=1/k_B T$) for weak and strong backscattering case (correspond to the full line and the dashed line). The insert is the ratio of non-equilibrium contribution in dephasing rate between the arbitrary screening and strong screening multiplied by $(\alpha/d)^2$ as a function of $d/\lambda_s$.}
\label{fig3}
\end{figure}
In Fig. \ref{fig3}, the dependence of the non-equilibrium contribution of the dephasing rate on the QPC bias voltage is plotted for several temperatures. In the case of strong BS, the dephasing rate increases when the bias $eV$ increases. When the temperature is low enough ($1/\beta\ll eV$), the dephasing rate saturates. In the case of high temperatures ($1/\beta > eV$), the dephasing rate also increases when $eV$ increases, but it increases from a finite value (not shown), which is proportional to the temperature. Things are quite different at weak BS. At high temperatures, the dephasing rate decreases  
when we increase $eV$: this behavior is symptomatic of current and noise characteristic in a Luttinger liquid. In the low temperature case $1/\beta\ll eV$, for small $eV$, the lower the temperature, the bigger the dephasing rate and the faster it decreases when we increase $eV$. At $T=0$, the dephasing rate is ``infinite'' at $eV=0$. This Luttinger liquid behavior is in sharp contrast with the result of Ref. \onlinecite{levinson_euro39}. 

The charge fluctuations are directly related to the current fluctuations along the edges. 
The fluctuations of the currents along the edges are also identical to the fluctuations of the tunneling current. The tunneling current fluctuations were computed non pertubatively using Bethe Ansatz techniques\cite{fendley_saleur}. We can therefore invoke current conservation at the point contact to derive a general formula for the decoherence rate, which describes the crossover from weak to strong BS\cite{fendleyPRL}: 
\begin{equation}
(\tau_{\varphi}^{-1})^{(2)}=\frac{e^3\tau_0^2}{ d^2}\frac{\nu}{1-\nu}(V G_{diff}-I)~,
\label{dephasing_bethe}\end{equation}
where $G_{diff}=\partial_V I$ is the differential conductance, $I$ is the current, defined 
in Refs. \onlinecite{fendleyPRL,fendleyPRB}. Eq. (\ref{dephasing_bethe}) allows to describe 
the crossover in the dephasing rate from the weak to the strong BS regime. 

Remarkably, for the weak and the strong BS regimes, 
it is possible to go beyond the strong screening limit, and one can 
compute Eq. (\ref{A2}) for an arbitrary Coulomb kernel $f(x)$: 
the triple time integral in the second order contribution to the dephasing rate
are computed analytically.
Further simplifications occur if $f(x)$ is even. The result can be displayed in terms of the 
ratio between the arbitrary screening dephasing rate and the strong screening dephasing rate (both non-equilibrium contributions): 
\begin{equation}
F\equiv\frac{(\tau_{\varphi}^{-1})^{(2)}}{(\tau_{\varphi}^{-1})^{(2)}_{\lambda_s\rightarrow \alpha}}=\frac{d^2}{(e\alpha)^2}\left[\int_{0}^{\infty}\!\!dx f(x)\right]^2~,
\label{arbitrary_screening}\end{equation}
where the integral is a function of $d/\lambda_s$, and we recall that $\alpha$ is the spatial cutoff. 
If the Coulomb interaction kernel $f(x)$ is chosen as
suggested before, the dephasing rate at arbitrary $\lambda_s$ 
has an analytical expression:  
$F=(\pi d/2\alpha)^2 [E_0(d/\lambda_s)+N_0(d/\lambda_s)]$, where $E_0(d/\lambda_s)$ and  $N_0(d/\lambda_s)$ are the Weber and 
the Neumann function\cite{abramowitz}, both of zero order. $F$ is plotted in the insert of Fig. \ref{fig3}, and $(\alpha/d)^2$ is taken to be
a small constant.
$F$ is infinite in the absence of screening, but in practical situations, the presence of metallic gates
always imposes a finite screening length. $F$ decreases with $d/\lambda_s$ and approaches $1$ when $\lambda_s$ is 
close to the spatial cutoff $\alpha$ (strong screening). The dephasing rate 
increases when the screening decreases.      

To summarize, we have established a general formula for the dephasing rate of a quantum dot located 
in the proximity of a fluctuating fractional edge current. 
In the case where screening is strong, we have shown that the dephasing rate is given by the tunneling current noise, 
regardless of the regime (weak or 
strong BS) which is considered. For weaker screening, the spatial 
dependence of the density-density correlation function has to be taken into account,
but we have shown explicitly that the long range nature of the Coulomb interaction can be 
included as a trivial multiplicative factor. We conjecture that in order to 
describe the crossover in the dephasing rate between weak and strong backscattering cases 
for arbitrary screening, it is sufficient to use
the strong screening crossover result of Eq. (\ref{dephasing_bethe}) and to 
insert it in Eq. (\ref{arbitrary_screening}). 
On the one hand, the fact that the dephasing rate decreases with increasing 
voltage can be reconciled with the fact that the charge noise is directly related to the BS 
current noise in the FQHE. There it is known, and seen experimentally, that 
when the  bias voltage dominates over the temperature, both 
the tunneling current and noise bear a power law 
dependence $\sim V^{2\nu-1}$ with a negative exponent. 
On the other hand, the fact that at low temperatures, 
the dephasing rate for filling factors can be lower than that of the integer quantum Hall effect 
comes as a surprise, which is contained in 
the temperature/voltage crossover formula of Eq. (9). It is yet another consequence 
of chiral Luttinger liquid theory.

The present results could be tested with gated heterostructures as in Ref. \onlinecite{sprinzak}
(see Fig. 4a of this work),
provided that the electron mobility and the magnetic field are further increased in order 
to achieve the FQHE regime and provided that the quantum dot is placed next to the QPC
as in Fig. 1.

\end{document}